# A VOLUME AVERAGING APPROACH TO THE INTERFACIAL FORCE IN HOMOGENEOUS BUBBLY FLOWS WITH MONODISPERSE BUBBLES

Sergio A. Baz-Rodríguez[*]

Facultad de Ingeniería Química, Universidad Autónoma de Yucatán, Campus de Ciencias Exactas e Ingenierías, Periférico Norte Kilómetro 33.5, Tablaje Catastral 13615, Col. Chuburná de Hidalgo Inn, Mérida, Yucatán, C. P. 97203, México.

**ABSTRACT**

The mathematical up-scaling of gas-liquid bubbly flows was carried out under the framework of the volume averaging theory. A two-fluid model and its associated closure problem were deduced. The closure problem was solved for a case study: a liquid flow past an assembly of homogeneous and monodisperse distorted bubbles. To this end, a cylindrical unit cells with concentric bubbles was used as approach of a representative elementary volume. The following conditions were evaluated: $0 < Re \leq 675$, $0 < \varepsilon_g \leq 0.2$, $7.15 \times 10^{-9} \leq We \leq 3.96$ and $Mo = 2.55 \times 10^{-11}$. The interfacial force closure was analyzed in terms of the ratio between the interfacial force for bubbly flow and that existing for a single bubble with the equivalent $Re$ (in an unbounded liquid). It was found that, for creeping regime ($0.001 \leq Re \leq 0.1$), the ratio between drag coefficients increases as $\varepsilon_g$ increases independently of $Re$. For intermediate and inertial regimes ($1 \leq Re \leq 650$), the ratio also increases as $\varepsilon_g$ increases; however, two different trends were observed with respect to $Re$: the ratio decreases for the intermediate regime ($1 \leq Re \leq 50$) and increases for the inertial regime ($50 < Re \leq 650$). A transition point around $Re = 50$ was found, which could be related to the boundary layer formation. Similar trends with respect to existing data from direct numerical simulations of bubbly flows were obtained, showing that this approach represents a good alternative, in terms of computing efforts, to obtain closure laws for two-phase flow modeling.

**Keywords**: interfacial drag force, gas-liquid bubbly flow, two-fluid model, closure problem.

## 1. INTRODUCTION

Gas-liquid bubbly flows are commonly present in chemical and biochemical industries. Gases are often compressed and dispersed as bubbles through liquids in bubble column reactors and bioreactors, as well as in absorption and stripping columns. Also, frequently in the petroleum industry, gas-liquid flows are conveyed through pressurized pipelines. Nowadays, the ability to properly predict the behavior of transport processes in these systems is still a complex problem for physics and applied mathematics.

The predictions of velocity, pressure and volumetric fraction average fields of each phase in gas-liquid flows is often obtained from the solution of averaged transport equations [1,2]. These equations overcome the problem that imply the spatial discontinuities of the different phases and the displacement of the interfaces; they are continuous over the entire domain, but at the expense of the occurrence of closure terms. These latter involve filtered information due to the spatial smoothing applied by averaging procedures when the up-scaled equations are deduced. The definition of constitutive relations for the closure terms is essential prior to the employment of averaged models to predict two-phase flow behavior.

---

[*] Email-address: sergio.baz@correo.uady.mx

Regarding the definition of closure terms of averaged models for two-phase flows, it is common the employing of a practical approach based on the direct usage in the up-scale of mathematical solutions for the flow around a disperse-phase representative body. It is interpreted that all the dispersed bodies in the system undergo the phenomena governed by invariant constitutive equations [3]. This approach can be supported on the principle of material objectivity and implies frame-indifference [4,5]. The precise definition of closure parameters for this approach often implies an associated unit cell problem, complemented by empirics and/or numerical findings [3,5,6]. An alternative is the volume averaging method with closure (VAMC), proposed by Whitaker [2]. This method goes beyond the deduction of average equations, since it proposes a mathematical formalism to deduce a set of boundary-value problems associated to the closure terms [7].

The up-scaling of bubbly flows with a focus on to study how interacting bubbles affect the interfacial forces has been object of wide research interest. Early experimental works studied the effects of the gas volumetric fraction and the mean bubble size on the rise velocity (which is related to the drag coefficient) of bubble swarms [8–11]. Their observations indicated that the drag force for individual bubbles in a swarm is larger than the corresponding one for an isolated single bubble. Later on, some specific spatial configurations were experimentally studied in order to analyze the rising swarm velocity or bubble coalescence behavior in systems with pure liquids: vertical trains [12–15] and angular or lateral arranges of pair of bubbles [16,17]. These works identified the relevance of bubble wakes on the drag reduction for rising bubbles interacting in-line, as well as highlighted the existence of angular-dependent attractive or repulsive forces between bubbles. Complementarily, analytical [18–22] and numerical [23–26] works have studied the interaction forces between pairs of bubbles under inviscid and/or viscous flow assumptions.

Bubbles moving through inhomogeneous flows (as occurs inside a bubble swarms) experience more effects than drag; namely, inertial, added mass, lift and history forces [27]. However, the drag is one or two orders of magnitude larger than the additional minor force contributions; therefore, its study is very relevant in the prediction of bubbly flow behavior.

The definition of interfacial force closures for averaged mixture or two-fluid models under a frame-indifferent approach is commonly expressed in terms of a drag swarm coefficient [1,5,28,29]. This parameter, analytically deduced and/or fitted from numerical or experimental data, has been extensively employed for momentum transport prediction in bubbly flows [30].

In this work, the mathematical up-scaling of gas-liquid bubbly flows was analyzed by using the VAMC as up-scaling method, with a focus on the interfacial force term. A matching with a conventional approach to the closure issue, supported on the principle of material objectivity, was carried out in terms of a drag coefficient. A particular scheme of gas-liquid bubbly flow represented by flow past an assembly of monodisperse bubbles was solved to show the usefulness of the method.

## 2. PROBLEM STATEMENT

Let us consider the isothermal motion of a dispersed gas phase through a liquid away to the effects of walls. A zoomed representation of the bulk of this system is given in the right side of Figure 1 [7]. Both phases are Newtonian and the gas motion is characterized by very low Mach numbers so that it can be considered as pseudo-incompressible. In Figure 1, $L$ is a characteristic macroscopic length and $l_l$ and $l_g$ are characteristic lengths where significant local changes occur in the liquid and gas phases, respectively. Similarly, a pair of times, $t_g$ and $t_l$, can be defined as characteristic times of instantaneous variation for gas and liquid. The boundary-value problem describing the fluid flow at the local and instantaneous scale is the following:

$$\rho_l \left( \frac{\partial \mathbf{v}_l}{\partial t'} + \mathbf{v}_l \cdot \nabla' \mathbf{v}_l \right) = -\nabla' p_l + \nabla' \cdot \boldsymbol{\tau}_l + \rho_l \mathbf{g} \quad \text{at the } l\text{-phase}, \tag{1}$$

$$\nabla' \cdot \mathbf{v}_l = 0 \quad \text{at the } l\text{-phase}, \tag{2}$$

$$\rho_g \left( \frac{\partial \mathbf{v}_g}{\partial t'} + \mathbf{v}_g \cdot \nabla' \mathbf{v}_g \right) = -\nabla' p_g + \nabla' \cdot \boldsymbol{\tau}_g + \rho_g \mathbf{g} \quad \text{at the } g\text{-phase}, \tag{3}$$

$$\nabla' \cdot \mathbf{v}_g = 0 \quad \text{at the } g\text{-phase}, \tag{4}$$

$$\mathbf{n}_{gl} \cdot (\mathbf{v}_g - \mathbf{w}) = -\mathbf{n}_{lg} \cdot (\mathbf{v}_l - \mathbf{w}) = 0 \quad \text{at } A_{gl}, \tag{5}$$

$$\mathbf{n}_{gl} \cdot (-p_g \mathbf{I} + \boldsymbol{\tau}_g) + \mathbf{n}_{lg} \cdot (-p_l \mathbf{I} + \boldsymbol{\tau}_l) = 2H_{gl} \sigma \mathbf{n}_{gl} \quad \text{at } A_{gl}, \tag{6}$$

$$\mathbf{v}_g = \mathcal{F}(\mathbf{x}', t'), \; p_g = \mathcal{K}(\mathbf{x}', t') \quad \text{at } \mathcal{A}_{ge}, \tag{7}$$

$$\mathbf{v}_l = \mathcal{G}(\mathbf{x}', t'), \; p_l = \mathcal{L}(\mathbf{x}', t') \quad \text{at } \mathcal{A}_{le}, \tag{8}$$

$$\mathbf{v}_g = \mathcal{H}(\mathbf{x}'), \; p_g = \mathcal{M}(\mathbf{x}') \quad \text{when } t' = 0, \tag{9}$$

$$\mathbf{v}_l = \mathcal{J}(\mathbf{x}'), \; p_l = \mathcal{N}(\mathbf{x}') \quad \text{when } t' = 0. \tag{10}$$

Equations (1)-(4) describe the conservation of mass and linear momentum for the studied phases. In them, $\rho$ is the density, $\mathbf{v}$ is the velocity vector, $p$ is the pressure, $\boldsymbol{\tau}$ is the viscous stress tensor and $\mathbf{g}$ is the vector of gravity acceleration; the subscript $k = g, l$ in the variables indicate that they correspond to gas and liquid phases, respectively. Equations (5) and (6) are boundary conditions indicating lack of interfacial mass flux and the momentum jump condition, respectively [1,31]; in them, $\mathbf{w}$ is the local interface velocity and $H_{gl}$ is the mean curvature of interface from the gas phase and $\sigma$ is the interfacial pressure; this latter is assumed as constant over the interfaces. $\mathbf{n}$ is the normal unit vector to the interface, defined to point toward the gas phase if subscript is $lg$ and toward the liquid phase if subscript is $gl$.

Finally, Equations (7)-(10) represent boundary and initial conditions for the involved variables. In Equations (5)-(8), the interfacial area between phases is identified by $A_{gl}$, while $A_{le}$ and $A_{ge}$ represent the areas of entrances to the global system for the liquid and gas phases, respectively. The viscous stress tensor is defined (for incompressible Newtonian fluids) as follows:

$$\boldsymbol{\tau}_k = \mu_k \left[ \nabla \mathbf{v}_k + (\nabla \mathbf{v}_k)^T \right], \tag{11}$$

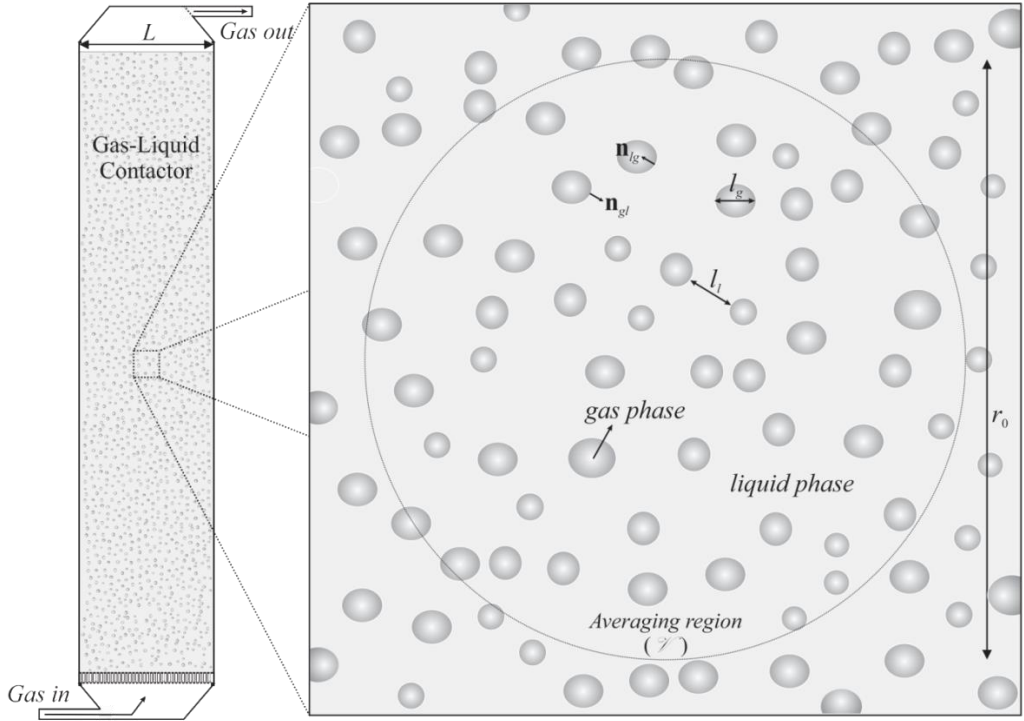

**Figure 1**. Geometrical representation of a gas-liquid bubbly flow [7].

where $\mu_k$ are the dynamic viscosity. If the gradients of velocity normal to interface are negligible (absence of normal oscillation at bubble interfaces), the continuity of momentum flux across interfaces [Equation (6)] is detachable in two decoupled contributions. These are: normal

$$p_g - p_l = 2H_{gl}\sigma, \qquad (12)$$

and tangencial to the interfaces

$$\mathbf{n}_{gl} \cdot \boldsymbol{\tau}_g \cdot \mathbf{t}_i + \mathbf{n}_{lg} \cdot \boldsymbol{\tau}_l \cdot \mathbf{t}_i = 0, \qquad (13)$$

where $\mathbf{t}_i$ represent any two unit vectors ($i = 1, 2$) which are tangential to the interface $gl$ and orthogonal between them. However, since the normal components of the stress tensor vanish under the abovementioned assumption, the following expression is also valid:

$$\mathbf{n}_{gl} \cdot \boldsymbol{\tau}_g + \mathbf{n}_{lg} \cdot \boldsymbol{\tau}_l = 0. \qquad (14)$$

A spatial and temporal control region ($\mathscr{V}$ in Figure 1) is defined; the region volume is $V$, and its characteristics size and time are $r_0$ and $T$, respectively. Any point in the control region can be located by the vector $\mathbf{x}'$ at time $t'$, whereas its spatial and temporal centroid is located by the vector $\mathbf{x}$ and time $t$. Hence, the prime symbol of differential operators and independent variables in Equations (1)-(10) indicate that they correspond to local and instantaneous description of the flow system. From the centroid of control region as origin, every point inside it can be located by a $\mathbf{y}$ vector, i.e., $\mathbf{x}' = \mathbf{x} + \mathbf{y}$. Similarly, every instant into the time period ($T$) from the central time can be located by a time difference $\theta$ ($t' = t + \theta$).

## 3. METHODS

### 3.1. Mathematical up-scaling

The VAMC [2] was applied to a general form of the governing equations given in the problem statement [Equations (1)-(4)]. For this, a set of operators and theorems are useful. Let us consider $\bullet_k = \bullet_k(\mathbf{x}',t')$ as any physical magnitude associated to the phase $k$, at the local and instantaneous scale of the system under study. From this, the following operators and theorems are defined [32]:
Spatiotemporal average:

$$\langle \bullet_k \rangle (\mathbf{x},t) = \frac{1}{TV} \int_{t-T/2}^{t+T/2} \int_{V_k(\mathbf{x},t)} \bullet_k(\mathbf{x}',t') dV'dt'. \tag{15}$$

Phase-intrinsic spatiotemporal average:

$$\langle \bullet_k \rangle^k (\mathbf{x},t) = \frac{1}{TV_k(\mathbf{x},t)} \int_{t-T/2}^{t+T/2} \int_{V_k(\mathbf{x},t)} \bullet_k(\mathbf{x}',t') dV'dt'. \tag{16}$$

Relation between averages [Equation (15) and (16)]:

$$\langle \bullet_k \rangle (\mathbf{x},t) = \varepsilon_k (\mathbf{x},t) \left[ \langle \bullet_k \rangle^k (\mathbf{x},t) \right], \tag{17}$$

where

$$\varepsilon_k (\mathbf{x},t) = \frac{V_k(\mathbf{x},t)}{V}. \tag{18}$$

Phase-intrinsic surface-temporal average (at the interface $km$):

$$\langle \bullet_k \rangle_{km} (\mathbf{x},t) = \frac{1}{TV_k(\mathbf{x},t)} \int_{t-T/2}^{t+T/2} \int_{A_{km}(\mathbf{x},t)} \bullet_k(\mathbf{x}',t') dA'dt'. \tag{19}$$

Space-time averaging theorem for a physical magnitude

$$\langle \nabla \cdot \bullet_k \rangle (\mathbf{x},t) = \nabla \cdot \langle \bullet_k \rangle (\mathbf{x},t) + \frac{1}{TV} \int_{t-T/2}^{t+T/2} \int_{A_{km}(\mathbf{x},t)} \mathbf{n}_{km} \cdot \bullet_k(\mathbf{x}',t') dA dt'. \tag{20}$$

Generalized theorem of transport:

$$\left\langle \frac{\partial \bullet_k}{\partial t} \right\rangle (\mathbf{x},t) = \frac{\partial \left[ \langle \bullet_k \rangle (\mathbf{x},t) \right]}{\partial t} - \frac{1}{TV} \int_{t-T/2}^{t+T/2} \int_{A_{km}(\mathbf{x},t)} \mathbf{n}_{km} \cdot \bullet_k(\mathbf{x}',t') \mathbf{w} dA dt'. \tag{21}$$

Roughly, the VAMC starts with the applying of operators and theorems to Equations (1)-(4) over the control region. After, decompositions of local variables (average plus the deviation around it):

$$\bullet_k = \langle \bullet_k \rangle^k + \tilde{\bullet}_k \tag{22}$$

are substituted [33,34]; if interest, reasonable scale constrains are assumed. The closure terms correspond to those including deviation variables. By deducing a boundary-value problem for deviation variables subject to scale disparity constrains, the closure terms can be solved in terms of closure variables. Details of the procedure and its development are given in Appendixes A and B.

**3.2. Drag coefficient for bubbly flows**

A practical approach based on the generalization of mathematical solutions for flow around a representative bubble often is employed to define an interfacial force [1]:

$$\mathbf{M}_{lg} = -\varepsilon_g c_D \frac{3}{4} \frac{\rho_l}{d_B} \left| \langle \mathbf{v}_g \rangle^g - \langle \mathbf{v}_l \rangle^l \right| \left( \langle \mathbf{v}_g \rangle^g - \langle \mathbf{v}_l \rangle^l \right), \tag{23}$$

where $c_D$ is the average drag coefficient and $d_B$ is the mean bubble diameter. Equation (23) was matched with the interfacial force closure obtained from VAMC in order to provide a way for the drag coefficient calculation.

**3.3. Case study**

An assessment of the VAMC predictions with respect to reported data was carried out for a case study. A form of the closure problem was solved for liquid flow past an assembly of homogeneous and monodisperse distorted bubbles. The associated constraints for closure variables were identified and the physical implications for bubbly flows were discussed. It was a matter of interest matching the drag coefficient results with those obtained by using a practical approach given by Equation (23).

Closure problems in VAMC often are associated to representative and periodic domains for solving. In some cases, numerical or analytical solutions from unit cells have proved to be in good agreement with experiments and DNS results [35–37]. By considering a single bubble surrounded by moving liquid as the elementary unit of the case under study, the closure problem can be solved for cylindrical unit cells with a centered bubble. This unit cell geometry ensures vertical periodicity and it is a rough representation of the smallest element of an ensemble of bubbles. Numerical calculations were carried out using Comsol Multiphysics® 3.5a. The system was represented in a 2D geometry with axial symmetry and several types of meshing were included in order for the results to be independent from the numerical parameters; an adaptive mesh refinement scheme was useful to ensure grid independence. The operating conditions and properties were grouped and specified in terms of dimensionless numbers: particle Reynolds, defined as $Re = d_B \left| \langle \mathbf{v}_g \rangle^g - \langle \mathbf{v}_l \rangle^l \right| \rho_l / \mu_l$, with respect to one mean superficial velocity; Weber, defined as $We = d_B \left| \langle \mathbf{v}_g \rangle^g - \langle \mathbf{v}_l \rangle^l \right|^2 \rho_l / \sigma$; and Morton, defined as $Mo = g \mu_l^4 / \rho_l \sigma^3$. The following conditions were evaluated: $0 < Re \leq 675$; $0 < \varepsilon_s \leq 0.2$; $7.15 \times 10^{-9} \leq We \leq 3.96$; $Mo = 2.55 \times 10^{-11}$.

*3.3.1. Single bubble motion*

The predictions of the proposed model were compared with the drag coefficient of the asymptotic case $\varepsilon_g \to 0$ ($\varepsilon_g = 0.001$). This condition approaches the classical problem of flow past a single bubble [7]. For spherical bubbles rising in water, several well-known analytical equations were compared: creeping flow ($Re \ll 1$), Oseen correction ($Re < 1$) [38], boundary layer solution ($50 < Re < 300$) [39], inviscid flow ($Re > 200$) [40]; the correlation of Mei et al [41] which fit the entire range with spherical shape also was taken into account:

$$C_{D0} = \frac{16}{Re}\left\{1+\left[\frac{8}{Re}+0.5\left(1+3.315Re^{-0.5}\right)\right]^{-1}\right\}. \tag{24}$$

From $Re \approx 200$ the shape of bubbles begins to have some deviation of spherical form [27,42]. A shape ratio ($X$, defined as the ratio between the major and minor axis of oblate spheroid) characterizes this evolution against hydrodynamic parameters and the medium properties. Particularly, the surface tension plays a crucial role on distortion. Because of this, dimensionless groups including surface tension are useful for characterizing the dependency of shape on hydrodynamic parameters and properties. Legendre et al. [43] developed the following correlation which relates pertinent dimensionless groups against the shape ratio:

$$X = \frac{1}{1-\tfrac{9}{64}We\left(1+0.2Mo^{1/10}We\right)^{-1}}. \tag{25}$$

Experimental works also have found relationship between dimensionless groups. Rastello et al. [44] proposed the following correlation:

$$Re = 2.05We^{2/3}Mo^{-1/5}. \tag{26}$$

The Equations (25) and (26) were used for describing the gradual change of shape distortion with respect to $Re$ [7]. Thus, the asymptotic case representing flow passing an isolated single bubble was solved for $0 < Re < 675$. This range is associated to rectilinear paths in the motion of single bubbles [45,46]. The results were compared with the analytical solution of Moore [47] for distorted single bubbles:

$$C_{D0} = \frac{48}{Re}G(X)\left[1+\frac{H(X)}{Re^{0.5}}\right] \tag{27}$$

where $H(X)$ and $G(X)$ are factors dependent on geometry; they are defined in the Moore's paper.

*3.3.2. Assembly of multiple bubbles*

An assessment of the model predictions for homogeneous and monodisperse bubbly flows was carried out. For this, the ratio between magnitudes of interfacial force for bubbly flow [Equation (23)] and that for a single bubble with the equivalent $Re$ in an unbounded liquid was calculated as follows:

$$\frac{c_D}{c_{D0}} = \frac{|\mathbf{M}_{lg}|}{|\mathbf{M}_{lg0}|}, \tag{28}$$

where $|\mathbf{M}_{lg0}|$ is the magnitude of the interfacial force closure for unbounded flow around a spherical particle; the Equation (27) was employed for the drag of an isolated spherical bubble. The ratio between forces [Equation (28)] also is equivalent to the correction function for drag coefficient [$f(\varepsilon_g,\ldots)$] reported for bubbly flows [48]:

$$\frac{c_D}{c_{D0}} = f\left(\varepsilon_g,\ldots\right). \tag{29}$$

The correction function often has been treated as a function only of $\varepsilon_g$; in this sense, the correlations of Bridge et al. [8], Lockett and Kirkpatrick [11], Rusche e Issa [48] and Simonnet et al. [49] can be stand out. On the other side, some correlations from DNS report co-dependence on dimensionless numbers:

Roghair et al. [50], with respect to $Eo$,

$$f\left(\varepsilon_g, Eo\right) = \left(1-\varepsilon_g\right)\left[1+\left(\frac{18}{Eo}\right)\varepsilon_g\right], \tag{30}$$

and Gillissen et al. [51], with respect to $Re$ and $We$,

$$f\left(\varepsilon_g, Re, We\right) = \frac{16}{Re_{eff} c_{D0}}\left\{1+\left[\frac{8}{Re_{eff}} + 0.5\left(1+3.315 Re_{eff}^{-0.5}\right)\right]^{-1}\right\}\left[1-\left(\frac{We}{4}\right)^{1.16}\right]^{-0.92}, \tag{31}$$

where

$$Re_{eff} = \frac{1}{\dfrac{1}{Re\left(1-0.6\varepsilon_g^{1/3}\right)} + 0.13\varepsilon_g}. \tag{32}$$

The correction factor for the case study from the deduced model and the above correlations were calculated. A correlation for the data obtained in this work was fitted in terms of $\varepsilon_g$ and $Re$.

## 4. RESULTS AND DISCUSSION

### 4.1. Averaged closed two-fluid model

Once applied the averaging procedure on local and instantaneous governing equations [Equations (1)-(4)], the following set of equations is obtained:

$$\nabla \cdot \langle \mathbf{v}_l \rangle^l = -\langle \mathbf{v}_l \rangle^l \cdot \frac{\nabla \varepsilon_l}{\varepsilon_l} - \frac{1}{\varepsilon_l}\frac{\partial \varepsilon_l}{\partial t}, \tag{33}$$

$$\rho_l\left(\frac{\partial \langle \mathbf{v}_l \rangle^l}{\partial t} + \langle \mathbf{v}_l \rangle^l \cdot \nabla \langle \mathbf{v}_l \rangle^l\right) = -\nabla \langle p_l \rangle^l + \mu_l \nabla^2 \langle \mathbf{v}_l \rangle^l + \rho_l \mathbf{g} + \frac{\nabla \varepsilon_l}{\varepsilon_l}\langle \Delta p_l \rangle_{lg}$$

$$+ \frac{\mathbf{M}_{lg}}{\varepsilon_l} + \frac{\nabla \cdot \left(\varepsilon_l \boldsymbol{\tau}_l^{turb}\right)}{\varepsilon_l}, \tag{34}$$

$$\nabla \cdot \langle \mathbf{v}_g \rangle^g = -\langle \mathbf{v}_g \rangle^g \cdot \frac{\nabla \varepsilon_g}{\varepsilon_g} - \frac{1}{\varepsilon_g} \frac{\partial \varepsilon_g}{\partial t}, \tag{35}$$

$$\rho_g \left( \frac{\partial \langle \mathbf{v}_g \rangle^g}{\partial t} + \langle \mathbf{v}_g \rangle^g \cdot \nabla \langle \mathbf{v}_g \rangle^g \right) = -\nabla \langle p_g \rangle^g + \mu_g \nabla^2 \langle \mathbf{v}_g \rangle^g + \rho_g \mathbf{g} + \frac{\nabla \varepsilon_g}{\varepsilon_g} \langle \Delta p_g \rangle_{gl}$$
$$+ \frac{\mathbf{M}_{gl}}{\varepsilon_g} + \frac{\nabla \cdot \left( \varepsilon_g \boldsymbol{\tau}_g^{turb} \right)}{\varepsilon_g}, \tag{36}$$

$$\langle \Delta p_l \rangle_{lg} = \langle p_l \rangle_{lg} - \langle p_l \rangle^l, \tag{37}$$

$$\langle \Delta p_g \rangle_{gl} = \langle p_g \rangle_{gl} - \langle p_g \rangle^g, \tag{38}$$

$$\langle p_g \rangle_{gl} - \langle p_l \rangle_{gl} = 2 \langle H_{gl} \rangle_{gl} \sigma. \tag{39}$$

The terms $\mathbf{M}_{gl}$ and $\mathbf{M}_{lg}$, and $\boldsymbol{\tau}_l^{turb}$ and $\boldsymbol{\tau}_g^{turb}$ in Equations (34) and (36) represent the interfacial force closures and the turbulent stress closures. The two-fluid model is subject to the following scale constrains:

$$l_l, l_g \ll r_0 \ll L, \quad t_l, t_g \ll T \ll \Theta, \tag{40}$$

Following the method, a boundary-value problem for deviation variables was deduced. Also, its source terms were identified. Solution forms for the deviation variables in terms of closure variables and source terms were proposed and substituted in the problem. After this, the following problem for the closure variables arises:

$$\nabla \cdot \mathbf{A}_l = 0 \quad \text{at the } l\text{-phase}, \tag{41}$$

$$\rho_l \mathbf{v}_l \cdot \nabla \mathbf{A}_l = -\nabla \mathbf{a}_l + \mu_l \nabla^2 \mathbf{A}_l - \frac{1}{TV_l} \int_{t-T/2}^{t+T/2} \int_{A_{gl}(\mathbf{x},t)} \mathbf{n}_{lg} \cdot \left( -\mathbf{a}_l \mathbf{I} + \mu_l \nabla \mathbf{A}_l \right) dA \, dt' \quad \text{at the } l\text{-phase},$$
$$\tag{42}$$

$$\mathbf{n}_{lg} \cdot \mathbf{A}_l = \mathbf{n}_{lg} \cdot \mathbf{A}_g + \mathbf{n}_{lg} \quad \text{at } A_{gl}, \tag{43}$$

$$\mathbf{a}_l = \mathbf{a}_g \quad \text{at } A_{gl}, \tag{44}$$

$$\mathbf{n}_{lg} \cdot \left( \nabla \mathbf{A}_l + \nabla \mathbf{A}_l^T \right) \cdot \mathbf{t} = \mathbf{n}_{lg} \cdot \frac{\mu_g}{\mu_l} \left( \nabla \mathbf{A}_g + \nabla \mathbf{A}_g^T \right) \quad \text{at } A_{gl}, \tag{45}$$

$$\nabla \cdot \mathbf{A}_g = 0 \quad \text{at the } g\text{-phase}, \tag{46}$$

$$\rho_g \mathbf{v}_g \cdot \nabla \mathbf{A}_g = -\nabla \mathbf{a}_g + \mu_g \nabla^2 \mathbf{A}_g - \frac{1}{TV_g} \int_{t-T/2}^{t+T/2} \int_{A_{gl}(\mathbf{x},t)} \mathbf{n}_{gl} \cdot \left( -\mathbf{a}_g \mathbf{I} + \mu_g \nabla \mathbf{A}_g \right) dA \, dt' \quad \text{at the } g\text{-phase},$$
$$\tag{47}$$

$$\mathbf{a}_l(\mathbf{r}+\mathbf{l}_i) = \mathbf{a}_l(\mathbf{r}), \quad \mathbf{a}_g(\mathbf{r}+\mathbf{l}_i) = \mathbf{a}_g(\mathbf{r}), \quad i=1,2,3, \tag{48}$$

$$\mathbf{A}_l(\mathbf{r}+\mathbf{l}_i)=\mathbf{A}_l(\mathbf{r}), \quad \mathbf{A}_g(\mathbf{r}+\mathbf{l}_i)=\mathbf{A}_g(\mathbf{r}), \quad i=1,2,3, \tag{49}$$

$$\langle \mathbf{A}_l \rangle^l, \langle \mathbf{A}_g \rangle^g, \langle \mathbf{a}_l \rangle^l, \langle \mathbf{a}_g \rangle^g = 0. \tag{50}$$

Equations (48) and (49) indicate periodicity of the solution fields. $\mathbf{A}_l$, $\mathbf{A}_g$, $\mathbf{a}_l$ and $\mathbf{a}_g$ are closure variables which are related with deviation variables in terms of source terms as follows:

$$\tilde{\mathbf{v}}_l = \mathbf{A}_l \cdot \left( \langle \mathbf{v}_g \rangle^g - \langle \mathbf{v}_l \rangle^l \right), \tag{51}$$

$$\tilde{p}_l = \mathbf{a}_l \cdot \left( \langle \mathbf{v}_g \rangle^g - \langle \mathbf{v}_l \rangle^l \right), \tag{52}$$

$$\tilde{\mathbf{v}}_g = \mathbf{A}_g \cdot \left( \langle \mathbf{v}_g \rangle^g - \langle \mathbf{v}_l \rangle^l \right), \tag{53}$$

$$\tilde{p}_g = \mathbf{a}_g \cdot \left( \langle \mathbf{v}_g \rangle^g - \langle \mathbf{v}_l \rangle^l \right). \tag{54}$$

The closure terms of the averaged closed model are defined in term of closure variables as follows:

$$\mathbf{M}_{lg} = \left[ \frac{\varepsilon_l}{TV_l} \int_{t-T/2}^{t+T/2} \int_{A_{gl}(\mathbf{x},t)} \mathbf{n}_{lg} \cdot \left( -\mathbf{a}_l \mathbf{I} + \mu_l \nabla \mathbf{A}_l \right) dA\, dt' \right] \cdot \left( \langle \mathbf{v}_g \rangle^g - \langle \mathbf{v}_l \rangle^l \right), \tag{55}$$

$$\mathbf{M}_{gl} = \left[ \frac{\varepsilon_g}{TV_g} \int_{t-T/2}^{t+T/2} \int_{A_{gl}(\mathbf{x},t)} \mathbf{n}_{gl} \cdot \left( -\mathbf{a}_g \mathbf{I} + \mu_g \nabla \mathbf{A}_g \right) dA\, dt' \right] \cdot \left( \langle \mathbf{v}_g \rangle^g - \langle \mathbf{v}_l \rangle^l \right). \tag{56}$$

$$\boldsymbol{\tau}_l^{turb} = -\rho_l \left( \langle \mathbf{v}_g \rangle^g - \langle \mathbf{v}_l \rangle^l \right) \cdot \langle \mathbf{A}_l^T \mathbf{A}_l \rangle^l \cdot \left( \langle \mathbf{v}_g \rangle^g - \langle \mathbf{v}_l \rangle^l \right), \tag{57}$$

$$\boldsymbol{\tau}_g^{turb} = -\rho_g \left( \langle \mathbf{v}_g \rangle^g - \langle \mathbf{v}_l \rangle^l \right) \cdot \langle \mathbf{A}_g^T \mathbf{A}_g \rangle^g \cdot \left( \langle \mathbf{v}_g \rangle^g - \langle \mathbf{v}_l \rangle^l \right). \tag{58}$$

The detailed procedure for obtaining the two-fluid model [Equations (33)-(39)] and the closure problem [Equations (41)-(50)] are given in Appendix A and Appendix B, respectively.

### 4.2. Drag coefficient for bubbly flows

By matching Equation (55) with a conventional two-fluid models form of the interfacial force closure [Equation (23)], the following equation was obtained:

$$C_D = \frac{4}{3} \frac{\varepsilon_l d_B}{\varepsilon_g \rho_l \left| \langle \mathbf{v}_g \rangle^g - \langle \mathbf{v}_l \rangle^l \right|^2} \left\| \left[ \frac{1}{TV_l} \int_{t-T/2}^{t+T/2} \int_{A_{gl}(\mathbf{x},t)} \mathbf{n}_{lg} \cdot \left( -\mathbf{a}_l \mathbf{I} + \mu_l \nabla \mathbf{A}_l \right) dA\, dt' \right] \cdot \left( \langle \mathbf{v}_g \rangle^g - \langle \mathbf{v}_l \rangle^l \right) \right\|. \tag{59}$$

### 4.3. Case study

For liquid flow past a fixed assembly bubbles the closure problem is expressed in the following form:

$$\nabla \cdot \mathbf{A}_l = 0 \quad \text{at the } l\text{-phase}, \tag{60}$$

$$\rho_l \mathbf{v}_l \cdot \nabla \mathbf{A}_l = -\nabla \mathbf{a}_l + \mu_l \nabla^2 \mathbf{A}_l - \frac{1}{TV_l} \int_{t-T/2}^{t+T/2} \int_{A_{gl}(\mathbf{x},t)} \mathbf{n}_{lg} \cdot \left(-\mathbf{a}_l \mathbf{I} + \mu_l \nabla \mathbf{A}_l\right) dA\, dt' \quad \text{at the } l\text{-phase}, \quad (61)$$

$$\mathbf{n}_{lg} \cdot \mathbf{A}_l = \mathbf{n}_{lg} \quad \text{at } A_{gl}, \tag{62}$$

$$\mathbf{a}_l = \mathbf{0} \quad \text{at } A_{gl}, \tag{63}$$

$$\mathbf{n}_{lg} \cdot \left(\nabla \mathbf{A}_l + \nabla \mathbf{A}_l^T\right) \cdot \mathbf{t} = 0 \quad \text{at } A_{gl}, \tag{64}$$

$$\mathbf{a}_l(\mathbf{r} + \mathbf{l}_i) = \mathbf{a}_l(\mathbf{r}), \quad i = 1,2,3, \tag{65}$$

$$\mathbf{A}_l(\mathbf{r} + \mathbf{l}_i) = \mathbf{A}_l(\mathbf{r}), \quad i = 1,2,3, \tag{66}$$

$$\langle \mathbf{A}_l \rangle^l, \langle \mathbf{a}_l \rangle^l = 0. \tag{67}$$

Equations (60)-(67) imply decoupling between the closure variables of liquid and gas phases, which are subject to the following constraints:

$$\mathbf{A}_g \ll \mathbf{I}, \quad \mathbf{a}_g \ll \mathbf{1}. \tag{66}$$

The main implications of these constraints are: i) the liquid flows around dispersed bubbles with totally free slip surfaces and the shear stress across the interface is negligible (a common assumption since $\mu_l \gg \mu_g$, and in the absence of surface-active substances), ii) the velocity deviations of the gaseous phase ($\tilde{\mathbf{v}}_g$) around the velocity average are negligible if compared to those of the liquid phase, and iii) the pressure deviations of both phases are negligible.

*4.3.1. Single bubble motion*

The motion of single bubbles has been extensively studied [42,52], and its terminal velocity and drag are used as references for motion in bubble swarms [48,53]. The simplified model [Equations (60)-(67)] was reliable for predicting the well-known drag coefficient behavior for spherical (average relative error < 0.029; reference: [41]), oblate spheroidal (average relative error < 0.008; reference: [47]) and bubbles in the range $0 < Re < 675$ (see Figure 2a,b). Consequently, the results for bubbly flows are shown in terms of the ratio between interfacial force for bubbly flow and that for an isolated single bubble with equivalent $Re$.

*4.3.2. Assembly of multiple bubbles*

For creeping regime in bubbly flows ($Re$ = 0.001, 0.01, 0.1), the ratio between drag coefficients increases as $\varepsilon_g$ increases but it is independent of $Re$ (see Figure 3). This is consistent with the expected behavior for only diffusive transport processes, where the microstructure completely defines the values of film coefficients [6,54]. On the other side, as $\varepsilon_g$ increases, $c_D/c_{D0}$ increases due to the confinement of vorticity fields of the closure variable in even more reduced spaces into the unit cells. Consistently, the increasing of $c_D/c_{D0}$ against $\varepsilon_g$ is observed for all flow regimes.

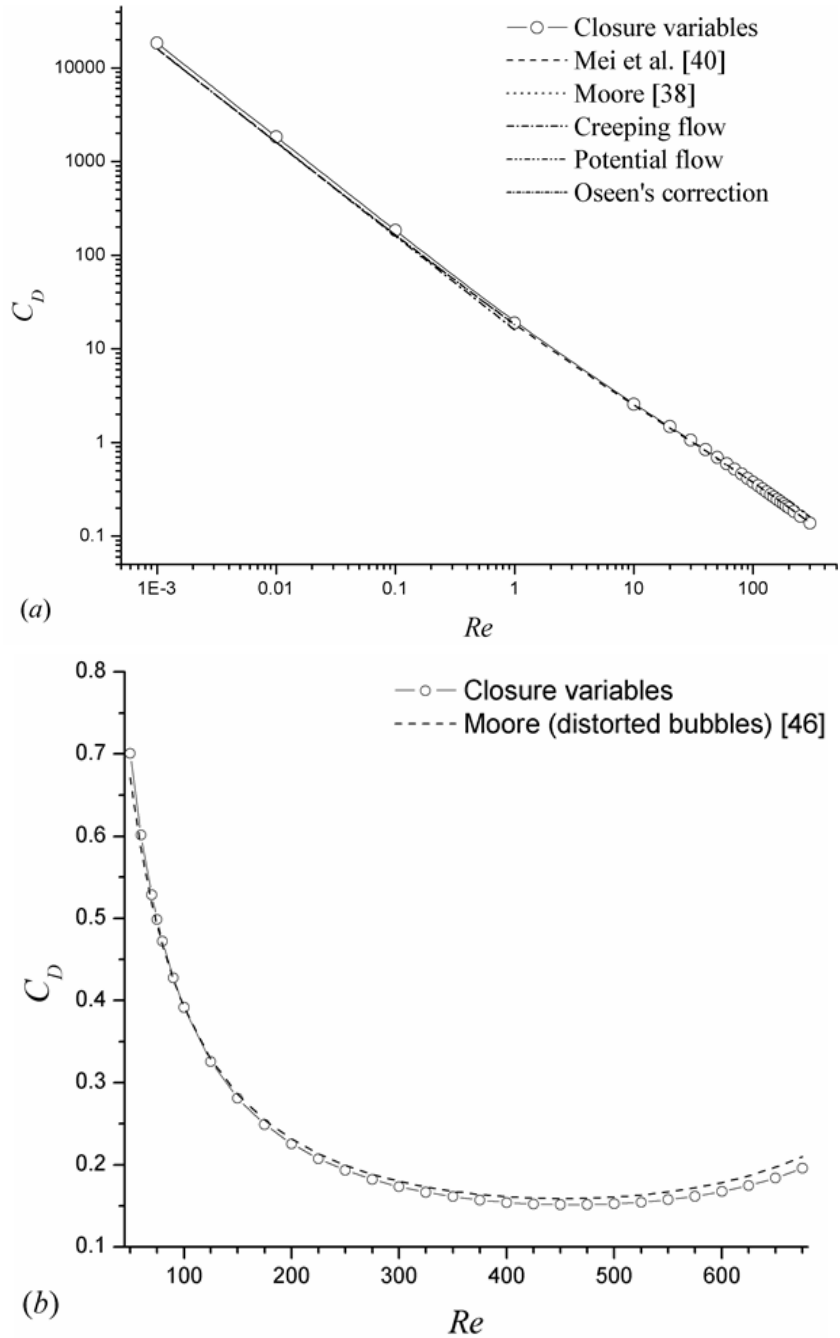

**Figure 2**. Drag coefficient for an isolated single bubble rising in an unbounded liquid: a) spherical bubble, and b) distorted bubble regime.

For low-to-moderate flow regime ($1 \leq Re \leq 50$), the ratio $c_D/c_{D0}$ decreases as $Re$ increases (see Figure 3). This range spans a transitional region between creeping and boundary layer flow regimes around bubbles. For isolated bubbles, there no exist analytical solutions for such transitional region [55] because the diffusive and advective contributions have similar orders of magnitude. The reduction in the ratio as $Re$ increases in such transitional regime is due to the increasing inertia and the partial advection downstream of generated vorticity in the closure variable field. As $Re$ is approaching to 50, the decreasing trend against the dimensionless group concludes and gives way to an opposite increasing behavior. Real bubbly flows with such local conditions often imply very small bubbles (e.g., $d_B$ < 0.7 mm, for air in water) [56] and significantly low volumetric gas fraction, which dampens the effect of interaction between bubbles. However, for closely interacting pairs of bubbles, it has been identified a transition point around $Re$ = 50 [57]. This value must be a transitional point between sufficiently large viscous effects and the irrotational mechanism which governs when boundary layers confine the vorticity near the bubble surface [24].

For a more inertial regime (50 ≤ $Re$ ≤ 650), the ratio $c_D/c_{D0}$ increases as $Re$ increases (see Figure 4). The predictions of the deduced model have similar trends and values than predictions from correlations reported for gravity driven bubbly flows [8,11,48–51]. Early correlations for drag in bubble swarms only involve dependence with $\varepsilon_g$ but other works have also correlated the influence of inertia (through $Re_{Eff}$ and $We$ [51]) and buoyancy (through $Eo$ [50]). The whole range predictions of these correlations are graphed in Figure 4. By using VAMC as a first approach for this simplified case study a reasonable agreement with reported behavior from DNS is achieved. This method proved to be a good alternative, in terms of computing efforts, to obtain closure laws by decoupling the problem for the deviation variables from the problem of the average variables. Therefore, this method is useful for the analysis and definition of closures.

## 5. CONCLUSIONS

From results of this work, the following conclusions stand out:
- The averaged two-fluid model for bubbly gas-liquid flows deduced by the VAMC involved an associated closure problem for the interfacial force in terms of a boundary-value problem for closure variables. This viewpoint enriches the state of the art as a useful alternative for obtaining constitutive relations for closure terms in two-fluid models.
- The interfacial force closure of bubbly flows, often expressed as the drag per area of a representative bubble directly up-scaled over the total specific area/volume, was related to the closure obtained via the VAMC. Thus, both approaches were matched.
- The closure problem was solved for a case study: liquid flow past an assembly of homogeneous and monodisperse distorted bubbles, using cylindrical unit cells with concentric bubbles as representative elementary volume ( 0 < $Re$ ≤ 675; 0 < $\varepsilon_s$ ≤ 0.2; 7.15 x $10^{-9}$ ≤ $We$ ≤ 3.96; $Mo$ = 2.55 x $10^{-11}$). The interfacial force closure was calculated as the ratio between the interfacial force from the closure problem solution and that for an isolated single bubble with the equivalent $Re$.

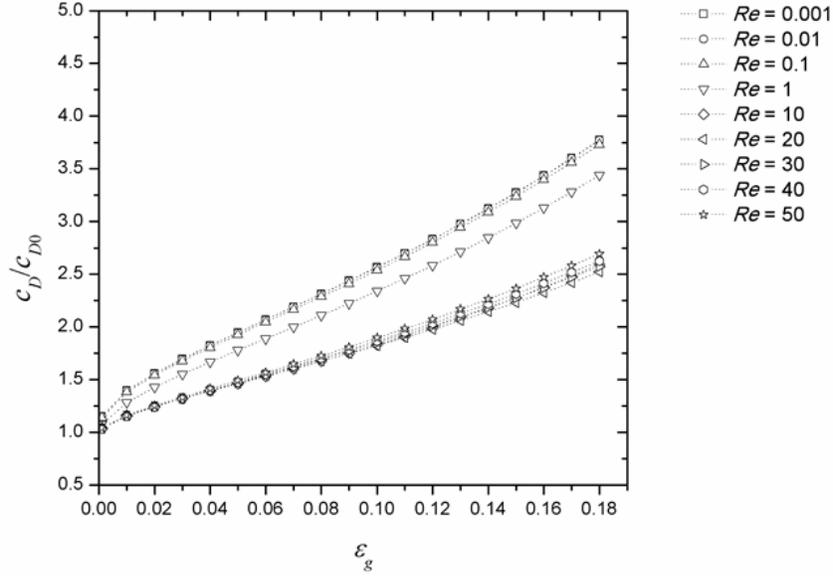

**Figure 3**. Normalized drag coefficient as a function of the gas volumetric fraction for low-to-moderate particle Reynolds numbers.

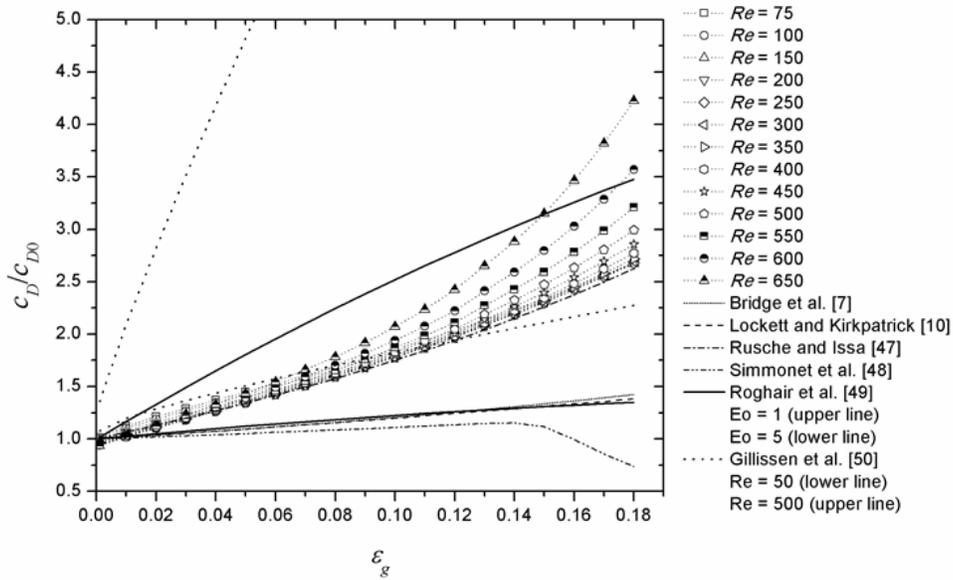

**Figure 4**. Normalized drag coefficient as a function of the gas volumetric fraction for large particle Reynolds numbers.

- The ratio $c_D/c_{D0}$ had values larger than 1 and increases as long as $\varepsilon_g$ did it. This implies an increase in the slip velocity as the gas phase in the system is more concentrated. Regarding the effect of $Re$, the ratio: i) was independent of the dimensionless group for creeping regime (0.001 ≤ $Re$ ≤ 0.1), ii) decreased with respect to $Re$ for intermediate regime (1 ≤ $Re$ ≤ 50). and iii) increased for inertial regime (50 < $Re$ ≤ 650). It was found, a transition point around $Re$ = 50, which could be related to the boundary layer formation and the irrotational behavior preeminence as governing mechanism when inertial effects increased.

**ACKNOWLEDGEMENTS**

The authors highly acknowledge the Consejo Nacional de Ciencia y Tecnología (CONACyT), México, for financial support through Grants 243423 and 316824.

**APPENDIX A. AVERAGING PROCEDURE AND THE TWO FLUID MODEL**

A general form of the boundary value problem describing the fluid flow at the local and instantaneous scale is the following:

$$\frac{\partial (\rho_l \mathbf{v}_l)}{\partial t'} + \nabla' \cdot (\rho_l \mathbf{v}_l \mathbf{v}_l) = -\nabla' p_l + \nabla' \cdot \boldsymbol{\tau}_l + \rho_l \mathbf{g} \quad \text{at the } l\text{-phase,} \tag{A.1}$$

$$\frac{\partial \rho_l}{\partial t'} + \nabla' \cdot (\rho_l \mathbf{v}_l) = 0 \quad \text{at the } l\text{-phase,} \tag{A.2}$$

$$\frac{\partial (\rho_g \mathbf{v}_g)}{\partial t'} + \nabla' \cdot (\rho_g \mathbf{v}_g \mathbf{v}_g) = -\nabla' p_g + \nabla' \cdot \boldsymbol{\tau}_g + \rho_g \mathbf{g} \quad \text{at the } g\text{-phase,} \tag{A.3}$$

$$\frac{\partial \rho_g}{\partial t'} + \nabla' \cdot (\rho_g \mathbf{v}_g) = 0 \quad \text{at the } g\text{-phase,} \tag{A.4}$$

$$\mathbf{n}_{gl} \cdot (\mathbf{v}_g - \mathbf{w}) = -\mathbf{n}_{lg} \cdot (\mathbf{v}_l - \mathbf{w}) = 0 \quad \text{at } A_{gl}, \tag{A.5}$$

$$\mathbf{n}_{gl} \cdot (-p_g \mathbf{I} + \boldsymbol{\tau}_g) + \mathbf{n}_{lg} \cdot (-p_l \mathbf{I} + \boldsymbol{\tau}_l) = 2H_{gl} \sigma \mathbf{n}_{gl} \quad \text{at } A_{gl}, \tag{A.6}$$

Equations (1)-(6) directly arise from this set of Equations by considering incompressible gas and liquid flows. By applying the averaging theorems [Equations (20) and (21)] to the continuity equation of the liquid phase [Equation (A.2)], we have:

$$\frac{\partial \langle \rho_l \rangle}{\partial t} + \nabla \cdot \langle \rho_l \mathbf{v}_l \rangle + \frac{1}{TV} \int_{t-T/2}^{t+T/2} \int_{A_{gl}(\mathbf{x},t)} \rho_l \mathbf{n}_{lg} \cdot (\mathbf{v}_l - \mathbf{w}) \, dA \, dt' = 0. \quad (A.7)$$

The last term of the Equation (A.7) vanishes due to the condition of non-existent interfacial mass flux [Equation (A.5)]:

$$\frac{\partial \langle \rho_l \rangle}{\partial t} + \nabla \cdot \langle \rho_l \mathbf{v}_l \rangle = 0. \quad (A.8)$$

By introducing the phase-intrinsic spatiotemporal average [Equation (16)], we obtain:

$$\frac{\partial \langle \rho_l \rangle}{\partial t} + \nabla \cdot \left( \varepsilon_l \langle \rho_l \mathbf{v}_l \rangle^l \right) = 0. \quad (A.9)$$

For convenience, a mass-weighted phase-intrinsic average is introduced for the velocity in the following form:

$$\{\mathbf{v}_k\}^k = \frac{\langle \rho_k \mathbf{v}_k \rangle^k}{\langle \rho_k \rangle^k}. \quad (A.10)$$

By substituting this weighted average in the Equation (A.9), we have:

$$\frac{\partial \left( \varepsilon_l \langle \rho_l \rangle^l \right)}{\partial t} + \nabla \cdot \left( \varepsilon_l \langle \rho_l \rangle^l \{\mathbf{v}_l\}^l \right) = 0. \quad (A.11)$$

For non-compressible liquids, the averaged mass conservation of the liquid phase is the following:

$$\frac{\partial \varepsilon_l}{\partial t} + \nabla \cdot \left( \varepsilon_l \langle \mathbf{v}_l \rangle^l \right) = 0. \quad (A.12)$$

By a similar procedure, the following averaged equation is obtained for the mass conservation in the gas phase:

$$\frac{\partial \varepsilon_g}{\partial t} + \nabla \cdot \left( \varepsilon_g \langle \mathbf{v}_g \rangle^g \right) = 0. \quad (A.13)$$

The Equations (33) and (35) directly arise from the development of the Equations (A.12) and (A.13).

By applying the averaging theorems [Equations (20) and (21)] to the equation for the momentum balance of the liquid phase [Equation (A.1)], we have:

$$\frac{\partial \langle \rho_l \mathbf{v}_l \rangle}{\partial t} + \nabla \cdot \langle \rho_l \mathbf{v}_l \mathbf{v}_l \rangle + \frac{1}{TV} \int_{t-T/2}^{t+T/2} \int_{A_{gl}(\mathbf{x},t)} \rho_l \mathbf{n}_{lg} \cdot \mathbf{v}_l (\mathbf{v}_l - \mathbf{w}) \, dA \, dt' = -\nabla \langle p_l \rangle + \nabla \cdot \langle \boldsymbol{\tau}_l \rangle$$
$$+ \langle \rho_l \rangle \mathbf{g} + \frac{1}{TV} \int_{t-T/2}^{t+T/2} \int_{A_{gl}(\mathbf{x},t)} \mathbf{n}_{lg} \cdot (-p_l \mathbf{I} + \boldsymbol{\tau}_l) \, dA \, dt' \quad (A.14)$$

Again, by the condition of non-existent interfacial mass flux [Equation (A.5)], the third term of the left side vanishes. Now, the following decomposition of the local and instantaneous shear tensor, according to the Equation (22),

$$\boldsymbol{\tau}_l = \langle \boldsymbol{\tau}_l \rangle^l + \tilde{\boldsymbol{\tau}}_l, \tag{A.15}$$

and the following decomposition for the pressure [34],

$$p_l = \langle p_l \rangle^l + \langle \Delta p_l \rangle_{lg} + \tilde{p}_l, \tag{A.16}$$

are introduced in the Equation (A.14). Then, we have:

$$\frac{\partial \langle \rho_l \mathbf{v}_l \rangle}{\partial t} + \nabla \cdot \langle \rho_l \mathbf{v}_l \mathbf{v}_l \rangle = -\nabla \langle p_l \rangle + \nabla \cdot \langle \boldsymbol{\tau}_l \rangle + \langle \rho_l \rangle \mathbf{g} + \frac{1}{TV} \int_{t-T/2}^{t+T/2} \int_{A_{gl}(\mathbf{x},t)} \mathbf{n}_{lg} \cdot \left( -\tilde{p}_l \mathbf{I} + \tilde{\boldsymbol{\tau}}_l \right) dA \, dt'$$

$$+ \frac{1}{TV} \int_{t-T/2}^{t+T/2} \int_{A_{gl}(\mathbf{x},t)} \mathbf{n}_{lg} \cdot \left( -\langle p_l \rangle^l \mathbf{I} - \langle \Delta p_l \rangle_{lg} \mathbf{I} + \langle \boldsymbol{\tau}_l \rangle^l \right) dA \, dt'.$$

$$\tag{A.17}$$

In the Equation (A.16):

$$\langle \Delta p_l \rangle_{lg} = \langle p_l \rangle_{lg} - \langle p_l \rangle^l, \tag{A.18}$$

and the following interfacial-intrinsic average is introduced:

$$\langle \bullet_k \rangle_{km} (\mathbf{x},t) = \frac{1}{TV_k(\mathbf{x},t)} \int_{t-T/2}^{t+T/2} \int_{A_{km}(\mathbf{x},t)} \bullet_k (\mathbf{x}',t') dV \, dt'. \tag{A.19}$$

The particular decomposition for the pressure arises because the difference between interfacial-intrinsic and phase-intrinsic average pressures deserves especial attention in bubbly flows. Cases involving flow separation in the wakes, at the rear point behind bubbles, or sudden expansion or shrinkage of interfaces are subject to significant differences between such averages [34,58].

The substituted decompositions involve local and instantaneous averages inside of integrals assigned to centroids of control volumes and integrating times. However, if

$$\langle p_g \rangle^g (\mathbf{x}',t') \approx \langle p_g \rangle^g (\mathbf{x},t), \tag{A.20}$$

$$\langle \Delta p_g \rangle_{gl} (\mathbf{x}',t') \approx \langle \Delta p_g \rangle_{gl} (\mathbf{x},t), \tag{A.21}$$

$$\langle \boldsymbol{\tau}_g \rangle^g (\mathbf{x}',t') \approx \langle \boldsymbol{\tau}_g \rangle^g (\mathbf{x},t), \tag{A.22}$$

the averages can be considered as constants into the integrals of Equation (A.17). For this, it is necessary that the following constraints to be fulfilled:

$$l_g \ll r_0 \ll L, \quad t_g \ll T \ll \theta. \tag{A.23}$$

Here, $L$ and $\theta$ are characteristic length and time in which significant changes of the average variables occur. Thus, the Equation (A.17) yields:

$$\frac{\partial \langle \rho_l \mathbf{v}_l \rangle}{\partial t} + \nabla \cdot \langle \rho_l \mathbf{v}_l \mathbf{v}_l \rangle = -\nabla \langle p_l \rangle + \nabla \cdot \langle \boldsymbol{\tau}_l \rangle + \langle \rho_l \rangle \mathbf{g} + \frac{1}{TV} \int_{t-T/2}^{t+T/2} \int_{A_{gl}(\mathbf{x},t)} \mathbf{n}_{lg} \cdot \left(-\tilde{p}_l \mathbf{I} + \tilde{\boldsymbol{\tau}}_l\right) dA\, dt'$$
$$+ \left( \frac{1}{TV} \int_{t-T/2}^{t+T/2} \int_{A_{gl}(\mathbf{x},t)} \mathbf{n}_{lg}\, dA\, dt' \right) \cdot \left(-\langle p_l \rangle^l \mathbf{I} - \langle \Delta p_l \rangle_{lg} \mathbf{I} + \langle \boldsymbol{\tau}_l \rangle^l \right).$$

(A.24)

From the space-time averaging theorem [Equation (20)]:

$$\frac{1}{TV} \int_{t-T/2}^{t+T/2} \int_{A_{km}(\mathbf{x},t)} \mathbf{n}_{km}\, dA\, dt' = -\nabla \langle 1 \rangle = -\nabla \varepsilon_k,$$

(A.25)

Thus, the Equation (A.24) yields:

$$\frac{\partial \langle \rho_l \mathbf{v}_l \rangle}{\partial t} + \nabla \cdot \langle \rho_l \mathbf{v}_l \mathbf{v}_l \rangle = -\nabla \langle p_l \rangle + \nabla \cdot \langle \boldsymbol{\tau}_l \rangle + \langle \rho_l \rangle \mathbf{g} + \frac{1}{TV} \int_{t-T/2}^{t+T/2} \int_{A_{gl}(\mathbf{x},t)} \mathbf{n}_{lg} \cdot \left(-\tilde{p}_l \mathbf{I} + \tilde{\boldsymbol{\tau}}_l\right) dA\, dt'$$
$$+ \nabla \varepsilon_k \cdot \left(\langle p_l \rangle^l \mathbf{I} + \langle \Delta p_l \rangle_{lg} \mathbf{I} - \langle \boldsymbol{\tau}_l \rangle^l \right).$$

(A.26)

By introducing the phase-intrinsic spatiotemporal average [Equation (16)] and the mass-weighted phase-intrinsic average [Equation (A.10)], we obtain:

$$\frac{\partial \left(\varepsilon_l \langle \rho_l \rangle^l \{\mathbf{v}_l\}^l\right)}{\partial t} + \nabla \cdot \left(\varepsilon_l \langle \rho_l \rangle^l \{\mathbf{v}_l \mathbf{v}_l\}^l\right) = -\nabla \left(\varepsilon_l \langle p_l \rangle^l\right) + \nabla \cdot \left(\varepsilon_l \langle \boldsymbol{\tau}_l \rangle^l\right) + \varepsilon_l \langle \rho_l \rangle^l \mathbf{g}$$
$$+ \frac{1}{TV} \int_{t-T/2}^{t+T/2} \int_{A_{gl}(\mathbf{x},t)} \mathbf{n}_{lg} \cdot \left(-\tilde{p}_l \mathbf{I} + \tilde{\boldsymbol{\tau}}_l\right) dA\, dt'$$
$$+ \nabla \varepsilon_l \cdot \left(\langle p_l \rangle^l \mathbf{I} + \langle \Delta p_l \rangle_{lg} \mathbf{I} - \langle \boldsymbol{\tau}_l \rangle^l \right).$$

(A.27)

Once developed the operations of the first two terms of the right side of Equation (A.27):

$$\frac{\partial \left(\varepsilon_l \langle \rho_l \rangle^l \{\mathbf{v}_l\}^l\right)}{\partial t} + \nabla \cdot \left(\varepsilon_l \langle \rho_l \rangle^l \{\mathbf{v}_l \mathbf{v}_l\}^l\right) = -\varepsilon_l \nabla \langle p_l \rangle^l + \varepsilon_l \nabla \cdot \langle \boldsymbol{\tau}_l \rangle^l + \varepsilon_l \langle \rho_l \rangle^l \mathbf{g} + \nabla \varepsilon_l \langle \Delta p_l \rangle_{lg}$$
$$+ \frac{1}{TV} \int_{t-T/2}^{t+T/2} \int_{A_{gl}(\mathbf{x},t)} \mathbf{n}_{lg} \cdot \left(-\tilde{p}_l \mathbf{I} + \tilde{\boldsymbol{\tau}}_l\right) dA\, dt'.$$

(A.28)

Now, the vectors in the dyadic tensor in the second term of the left side of the Equation (A.28) are substituted according to the following decomposition:

$$\mathbf{v}_k = \{\mathbf{v}_k\}^k + \tilde{\mathbf{v}}_k.$$

(A.29)

Taking into account the scale constraints given by the Equations (A.23), the Equation (A.28) yields:

$$\frac{\partial\left(\varepsilon_l\langle\rho_l\rangle^l\{\mathbf{v}_l\}^l\right)}{\partial t}+\nabla\cdot\left(\varepsilon_l\langle\rho_l\rangle^l\{\mathbf{v}_l\}^l\{\mathbf{v}_l\}^l\right)=-\varepsilon_l\nabla\langle p_l\rangle^l+\varepsilon_l\nabla\cdot\langle\boldsymbol{\tau}_l\rangle^l+\varepsilon_l\langle\rho_l\rangle^l\mathbf{g}+\nabla\varepsilon_l\langle\Delta p_l\rangle_{lg}$$
$$+\frac{1}{TV}\int_{t-T/2}^{t+T/2}\int_{A_{gl}(\mathbf{x},t)}\mathbf{n}_{lg}\cdot\left(-\tilde{p}_l\mathbf{I}+\tilde{\boldsymbol{\tau}}_l\right)dA\,dt'$$
$$-\nabla\cdot\left(\varepsilon_l\langle\rho_l\rangle^l\{\tilde{\mathbf{v}}_l\tilde{\mathbf{v}}_l\}^l\right).$$

(A.30)

From the Equation (A.30) arise the following closure terms:

$$\mathbf{M}_{lg}=\frac{1}{TV}\int_{t-T/2}^{t+T/2}\int_{A_{gl}(\mathbf{x},t)}\mathbf{n}_{lg}\cdot\left(-\tilde{p}_l\mathbf{I}+\tilde{\boldsymbol{\tau}}_l\right)dA\,dt', \quad (A.31)$$

$$\boldsymbol{\tau}_l^{turb}=-\langle\rho_l\rangle^l\{\tilde{\mathbf{v}}_l\tilde{\mathbf{v}}_l\}^l, \quad (A.32)$$

Being $\mathbf{M}_{lg}$ the interfacial force closure and $\boldsymbol{\tau}_l^{turb}$ the turbulent stresses [1,58]. By the averaged equation for mass conservation [Equation (A.11)] and incompressible flow, the Equation (A.30) yields:

$$\rho_l\left(\frac{\partial\langle\mathbf{v}_l\rangle^l}{\partial t}+\langle\mathbf{v}_l\rangle^l\cdot\nabla\langle\mathbf{v}_l\rangle^l\right)=-\nabla\langle p_l\rangle^l+\nabla\cdot\langle\boldsymbol{\tau}_l\rangle^l+\rho_l\mathbf{g}+\frac{\nabla\varepsilon_l}{\varepsilon_l}\langle\Delta p_l\rangle_{lg}$$
$$+\frac{\mathbf{M}_{lg}}{\varepsilon_l}+\frac{\nabla\cdot\left(\varepsilon_l\boldsymbol{\tau}_l^{turb}\right)}{\varepsilon_l},$$

(A.33)

By a similar procedure, the averaged momentum equation for the gas phase yields:

$$\rho_g\left(\frac{\partial\langle\mathbf{v}_g\rangle^g}{\partial t}+\langle\mathbf{v}_g\rangle^g\cdot\nabla\langle\mathbf{v}_g\rangle^g\right)=-\nabla\langle p_g\rangle^g+\nabla\cdot\langle\boldsymbol{\tau}_g\rangle^g+\rho_g\mathbf{g}+\frac{\nabla\varepsilon_g}{\varepsilon_g}\langle\Delta p_g\rangle_{gl}$$
$$+\frac{\mathbf{M}_{gl}}{\varepsilon_g}+\frac{\nabla\cdot\left(\varepsilon_g\boldsymbol{\tau}_g^{turb}\right)}{\varepsilon_g},$$

(A.34)

where:

$$\mathbf{M}_{gl}=\frac{1}{TV}\int_{t-T/2}^{t+T/2}\int_{A_{gl}(\mathbf{x},t)}\mathbf{n}_{gl}\cdot\left(-\tilde{p}_g\mathbf{I}+\tilde{\boldsymbol{\tau}}_g\right)dA\,dt', \quad (A.35)$$

$$\boldsymbol{\tau}_g^{turb}=-\langle\rho_g\rangle^g\{\tilde{\mathbf{v}}_g\tilde{\mathbf{v}}_g\}^g, \quad (A.36)$$

The Equations (34) and (36) directly arise from the development of the Equations (A.33) and (A.34) for Newtonian fluids.

## APPENDIX B. THE CLOSURE PROBLEM

The volume averaging method with closure takes advantage of the following mathematical relations:

$$\tilde{\bullet}_k = \bullet_k - \langle \bullet_k \rangle^k, \tag{B.1}$$

$$\nabla = \nabla', \frac{\partial}{\partial t} = \frac{\partial}{\partial t'}, \tag{B.2}$$

$$\mathcal{L}\tilde{\bullet}_k = \mathcal{L}\bullet_k - \mathcal{L}\langle \bullet_k \rangle^k. \tag{B.3}$$

In Equation (B.3), $\mathcal{L}$ represents the operators ( gradients, divergences, Laplacians and temporal derivatives) appearing in this work, which are linear. Consistently, by subtracting the averaged equations from the corresponding local and instantaneous equations, it is possible to obtain a set of equations delineating a problem for the deviation variables [2].

### B.1. From the conservation of mass
By subtracting Equation (33) from Equation (2), we obtain:

$$\nabla \cdot \tilde{\mathbf{v}}_l = \langle \mathbf{v}_l \rangle^l \frac{\nabla \varepsilon_l}{\varepsilon_l} + \frac{1}{\varepsilon_l} \frac{\partial \varepsilon_l}{\partial t}. \tag{B.4}$$

The orders of magnitude of each term in Equation (B.4) are given below:

$$O(\nabla \cdot \tilde{\mathbf{v}}_l) = \frac{\tilde{\mathbf{v}}_l}{l_{\tilde{\mathbf{v}}_l}}, \tag{B.5}$$

$$O\left(\langle \mathbf{v}_l \rangle^l \frac{\nabla \varepsilon_l}{\varepsilon_l}\right) = \frac{\langle \mathbf{v}_l \rangle^l}{L_{\varepsilon_l}}, \tag{B.6}$$

$$O\left(\frac{1}{\varepsilon_l} \frac{\partial \varepsilon_l}{\partial t}\right) = \frac{1}{t_{\varepsilon_l}}. \tag{B.7}$$

Here, $l_{\tilde{\mathbf{v}}_l}$ and $L_{\varepsilon_l}$ are characteristic lengths in which significant changes of $\tilde{\mathbf{v}}_l$ and $\langle \mathbf{v}_l \rangle^l$ occur, respectively, and $t_{\varepsilon_l}$ is the characteristic time in which significant changes of $\varepsilon_l$ occur. It is assumed that the differences between the deviation and the average velocities are in the same magnitude order. $L_{\varepsilon_l}$ is a length in the macroscale and $l_{\tilde{\mathbf{v}}_l}$ is a length in the microscale, i. e.,

$$L_{\varepsilon_l} \gg l_{\tilde{\mathbf{v}}_l}. \tag{B.8}$$

Also, the characteristic time for $\varepsilon_l$ (macroscopic variable) must be larger than the characteristic time of spatial variation of $\tilde{\mathbf{v}}_l$, i. e.,

$$t_{\varepsilon_l} \gg l_{\tilde{\mathbf{v}}_l} / \tilde{\mathbf{v}}_l, \tag{B.9}$$

Once taked into account the constraints given by Equations (B.8) and (B.9), the Equation (B.4) yields

$$\nabla \cdot \tilde{\mathbf{v}}_l = 0. \tag{B.10}$$

By subtracting Equation (35) from Equation (4) and taking into account similar scale constraints, we have:

$$\nabla \cdot \tilde{\mathbf{v}}_g = 0. \tag{B.11}$$

*B.2. From the conservation of momentum*

Now, by subtracting Equation (34) from Equation (1), we obtain:

$$\rho_l \left( \frac{\partial \tilde{\mathbf{v}}_l}{\partial t} + \mathbf{v}_l \cdot \nabla \mathbf{v}_l - \langle \mathbf{v}_l \rangle^l \cdot \nabla \langle \mathbf{v}_l \rangle^l \right) = -\nabla \tilde{p}_l - \nabla \langle \Delta p_l \rangle_{lg} + \nabla \cdot \tilde{\boldsymbol{\tau}}_l - \frac{\nabla \varepsilon_l}{\varepsilon_l} \langle \Delta p_l \rangle_{lg}$$

$$- \frac{\varepsilon_l}{TV} \int_{t-T/2}^{t+T/2} \int_{A_{gl}(\mathbf{x},t)} \mathbf{n}_{lg} \cdot \left( -\tilde{p}_l \mathbf{I} + \tilde{\boldsymbol{\tau}}_l \right) dA\, dt' + \frac{\rho_l}{\varepsilon_l} \nabla \cdot \left( \varepsilon_l \langle \tilde{\mathbf{v}}_l \tilde{\mathbf{v}}_l \rangle^l \right). \tag{B.12}$$

The spatial and temporal decomposition of the velocity, according to Equation (22) is now substituted in the second term of the left side of Equation (B.12):

$$\mathbf{v}_l \cdot \nabla \mathbf{v}_l = \tilde{\mathbf{v}}_l \cdot \nabla \tilde{\mathbf{v}}_l + \langle \mathbf{v}_l \rangle^l \cdot \nabla \tilde{\mathbf{v}}_l + \tilde{\mathbf{v}}_l \cdot \nabla \langle \mathbf{v}_l \rangle^l + \langle \mathbf{v}_l \rangle^l \cdot \nabla \langle \mathbf{v}_l \rangle^l. \tag{B.13}$$

Thus, the Equation (B.12) yields:

$$\rho_l \left( \frac{\partial \tilde{\mathbf{v}}_l}{\partial t} + \tilde{\mathbf{v}}_l \cdot \nabla \tilde{\mathbf{v}}_l + \langle \mathbf{v}_l \rangle^l \cdot \nabla \tilde{\mathbf{v}}_l + \tilde{\mathbf{v}}_l \cdot \nabla \langle \mathbf{v}_l \rangle^l \right) = -\nabla \tilde{p}_l - \nabla \langle \Delta p_l \rangle_{lg} + \nabla \cdot \tilde{\boldsymbol{\tau}}_l - \frac{\nabla \varepsilon_l}{\varepsilon_l} \langle \Delta p_l \rangle_{lg}$$

$$- \frac{\varepsilon_l}{TV} \int_{t-T/2}^{t+T/2} \int_{A_{gl}(\mathbf{x},t)} \mathbf{n}_{lg} \cdot \left( -\tilde{p}_l \mathbf{I} + \tilde{\boldsymbol{\tau}}_l \right) dA\, dt' + \frac{\rho_l}{\varepsilon_l} \nabla \cdot \left( \varepsilon_l \langle \tilde{\mathbf{v}}_l \tilde{\mathbf{v}}_l \rangle^l \right). \tag{B.14}$$

The orders of magnitude of the terms in left side of Equation (B.14) are given below:

$$O\left( \frac{\partial \tilde{\mathbf{v}}_l}{\partial t} \right) = \frac{\tilde{\mathbf{v}}_l}{t_{\tilde{\mathbf{v}}_l}}, \tag{B.15}$$

$$O\left( \tilde{\mathbf{v}}_l \cdot \nabla \tilde{\mathbf{v}}_l \right) = \frac{\tilde{\mathbf{v}}_l \tilde{\mathbf{v}}_l}{l_{\tilde{\mathbf{v}}_l}}, \tag{B.16}$$

$$O\left( \tilde{\mathbf{v}}_l \cdot \nabla \langle \mathbf{v}_l \rangle^l \right) = \frac{\tilde{\mathbf{v}}_l \langle \mathbf{v}_l \rangle^l}{L_{\langle \mathbf{v}_l \rangle}}, \tag{B.17}$$

$$O\left( \langle \mathbf{v}_l \rangle^l \cdot \nabla \tilde{\mathbf{v}}_l \right) = \frac{\tilde{\mathbf{v}}_l \langle \mathbf{v}_l \rangle^l}{l_{\tilde{\mathbf{v}}_l}}. \tag{B.18}$$

If the following scale constraints are satisfied:

$$L_{\langle \mathbf{v}_l \rangle} \gg l_{\tilde{\mathbf{v}}_l}, \quad t_{\tilde{\mathbf{v}}_l} \gg \frac{l_{\tilde{\mathbf{v}}_l}}{\tilde{\mathbf{v}}_l}, \tag{B.19}$$

the Equation (B.14) can be simplified to the following form:

$$\rho_l \mathbf{v}_l \cdot \nabla \tilde{\mathbf{v}}_l = -\nabla \tilde{p}_l - \nabla \langle \Delta p_l \rangle_{lg} + \nabla \cdot \tilde{\boldsymbol{\tau}}_l - \frac{\nabla \varepsilon_l}{\varepsilon_l} \langle \Delta p_l \rangle_{lg}$$
$$-\frac{\varepsilon_l}{TV} \int_{t-T/2}^{t+T/2} \int_{A_{gl}(\mathbf{x},t)} \mathbf{n}_{lg} \cdot \left(-\tilde{p}_l \mathbf{I} + \tilde{\boldsymbol{\tau}}_l\right) dA\, dt' + \frac{\rho_l}{\varepsilon_l} \nabla \cdot \left(\varepsilon_l \langle \tilde{\mathbf{v}}_l \tilde{\mathbf{v}}_l \rangle^l\right). \quad (B.20)$$

The scale constraints [Equations (B.19)] identify the differences between the magnitude orders of variables having relevant changes in the macroscale and the microscale. The right side of the Equation (B.20) is simplified under as indicated below:

$$\frac{\rho_l}{\varepsilon_l} \nabla \cdot \left(\varepsilon_l \langle \tilde{\mathbf{v}}_l \tilde{\mathbf{v}}_l \rangle^l\right) \ll \rho_l \tilde{\mathbf{v}}_l \cdot \nabla \tilde{\mathbf{v}}_l \quad \text{if } l_{\tilde{\mathbf{v}}_l} \ll L_{\tilde{\mathbf{v}}_l \tilde{\mathbf{v}}_l}, L_{\varepsilon_l}, \quad (B.21)$$

$$\nabla \langle \Delta p_l \rangle_{lg}, \frac{\nabla \varepsilon_l}{\varepsilon_l} \langle \Delta p_l \rangle_{lg} \ll \nabla \tilde{p}_l \quad \text{if } l_{pl} \ll L_{\Delta pl}, L_\varepsilon \text{ and } O\left(\langle \Delta p_l \rangle_{lg}\right) \approx O(\tilde{p}_l). \quad (B.22)$$

Finally, the Equation (B.20) yields:

$$\rho_l \mathbf{v}_l \cdot \nabla \tilde{\mathbf{v}}_l = -\nabla \tilde{p}_l + \nabla \cdot \tilde{\boldsymbol{\tau}}_l - \frac{\varepsilon_l}{TV} \int_{t-T/2}^{t+T/2} \int_{A_{gl}(\mathbf{x},t)} \mathbf{n}_{lg} \cdot \left(-\tilde{p}_l \mathbf{I} + \tilde{\boldsymbol{\tau}}_l\right) dA\, dt'. \quad (B.23)$$

Analogously, for the gas phase, we obtain:

$$\rho_g \left(\mathbf{v}_g \cdot \nabla \tilde{\mathbf{v}}_g\right) = -\nabla \tilde{p}_g + \nabla \cdot \tilde{\boldsymbol{\tau}}_g - \frac{1}{TV_g} \int_{t-T/2}^{t+T/2} \int_{A_{gl}(\mathbf{x},t)} \mathbf{n}_{gl} \cdot \left(-\tilde{p}_g \mathbf{I} + \tilde{\boldsymbol{\tau}}_g\right) dA\, dt'. \quad (B.24)$$

### B.3. From the boundary conditions

Now, the boundary conditions [Equations (5), (12) and (14)] are decomposed using the Equation (22) for the velocity vector and the Equation (A.16) for the pressure. The following equations arise:

$$\mathbf{n}_{lg} \cdot \tilde{\mathbf{v}}_l = \mathbf{n}_{lg} \cdot \tilde{\mathbf{v}}_g + \mathbf{n}_{lg} \cdot \left(\langle \mathbf{v}_g \rangle^g - \langle \mathbf{v}_l \rangle^l\right) \quad \text{at } A_{gl}, \quad (B.25)$$

$$\tilde{p}_l = \tilde{p}_g + \left(\langle p_g \rangle_{gl} - \langle p_l \rangle_{gl}\right) - 2H_{gl}\sigma \quad \text{at } A_{gl}, \quad (B.26)$$

$$\mathbf{n}_{lg} \cdot \tilde{\boldsymbol{\tau}}_l = \mathbf{n}_{lg} \cdot \tilde{\boldsymbol{\tau}}_g + \mathbf{n}_{lg} \cdot \left(\langle \boldsymbol{\tau}_g \rangle^g - \langle \boldsymbol{\tau}_l \rangle^l\right). \quad (B.27)$$

The surface-intrinsic averaging of the pressure jump boundary condition [Equation (12)] at constant surface tension yields:

$$\langle p_g \rangle_{gl} - \langle p_l \rangle_{gl} = 2\langle H_{gl} \rangle_{gl} \sigma. \quad (B.28)$$

By substituting Equation (B.28) in Equation (B.26), we have:

$$\tilde{p}_l = \tilde{p}_g + 2\left(\langle H_{gl} \rangle_{gl} - H_{gl}\right)\sigma \quad \text{at } A_{gl}. \quad (B.29)$$

In absence of significant oscillations of the shape of the bubbles, the mean curvature of interface remains as constant, then $\langle H_{gl} \rangle_{gl} \approx H_{gl}$ and the Equation (B.29) yields:

$$\tilde{p}_l = \tilde{p}_g \quad \text{at } A_{gl}. \tag{B.30}$$

Now, it is convenient substituting the shear stress tensors by constitutive relations in terms of field variables as velocity. Thus, the Equation (B.27) yields (Newtonian fluids):

$$\mathbf{n}_{lg} \cdot \left[ \nabla \tilde{\mathbf{v}}_l + \left( \nabla \tilde{\mathbf{v}}_l \right)^T \right] = \mathbf{n}_{lg} \cdot \frac{\mu_g}{\mu_l} \left[ \nabla \tilde{\mathbf{v}}_g + \left( \nabla \tilde{\mathbf{v}}_g \right)^T \right] + \mathbf{n}_{lg} \cdot \frac{\mu_g}{\mu_l} \left[ \nabla \langle \mathbf{v}_g \rangle^g + \left( \nabla \langle \mathbf{v}_g \rangle^g \right)^T \right]$$
$$- \mathbf{n}_{lg} \cdot \left[ \nabla \langle \mathbf{v}_l \rangle^l + \left( \nabla \langle \mathbf{v}_l \rangle^l \right)^T \right] \quad \text{at } A_{gl}. \tag{B.31}$$

It is assumed that the average velocities of each phase have similar or not very different magnitude orders than their respective deviation velocities. However, their changes take place in a larger characteristic length than that in which the deviation velocities change because the average velocities are of macroscopic nature. Thus,

$$l_{\tilde{\mathbf{v}}_k} \ll L_{\langle \mathbf{v}_k \rangle}, \tag{B.32}$$

and

$$\nabla \langle \mathbf{v}_k \rangle^k \ll \nabla \tilde{\mathbf{v}}_k, \quad k = l, g. \tag{B.33}$$

Accordingly, the Equation (B.31) yields

$$\mathbf{n}_{lg} \cdot \left[ \nabla \tilde{\mathbf{v}}_l + \left( \nabla \tilde{\mathbf{v}}_l \right)^T \right] = \mathbf{n}_{lg} \cdot \frac{\mu_g}{\mu_l} \left[ \nabla \tilde{\mathbf{v}}_g + \left( \nabla \tilde{\mathbf{v}}_g \right)^T \right] \quad \text{at } A_{gl}. \tag{B.34}$$

Finally, it is assumed that the gradients of the velocity deviations in the liquid phase either are in the same magnitude order or they are larger than those in the gas phase. Since $\mu_g \ll \mu_l$, the Equation (B.34) yields:

$$\mathbf{n}_{lg} \cdot \left[ \nabla \tilde{\mathbf{v}}_l + \left( \nabla \tilde{\mathbf{v}}_l \right)^T \right] = \mathbf{0} \quad \text{at } A_{gl}. \tag{B.35}$$

The Equations (B.25), (B.30) and (B.35) describe the boundary conditions for the deviation variables problem [Equations (B.10), (B.11), (B.23) and (B.24)].

*B.4. Relation between the interfacial force closures for liquid and gas*

Let us consider the definition of interfacial force closure from the liquid side [Equation (A.31). By substituting in such Equation the Equations (B.30) and (B.27), we have:

$$\mathbf{M}_{lg} = \frac{1}{TV} \int_{t-T/2}^{t+T/2} \int_{A_{gl}(\mathbf{x},t)} \mathbf{n}_{lg} \cdot \left( -\tilde{p}_g \mathbf{I} + \tilde{\boldsymbol{\tau}}_g \right) dA \, dt' + \frac{1}{TV} \int_{t-T/2}^{t+T/2} \int_{A_{gl}(\mathbf{x},t)} \mathbf{n}_{lg} \cdot \left( \langle \boldsymbol{\tau}_g \rangle^g - \langle \boldsymbol{\tau}_l \rangle^l \right) dA \, dt' \tag{B.36}$$

If the scale constraints given by Equation (40) are satisfied, i. e., the characteristic length of variations of the average variables is of some magnitude order than the deviation variables, the average terms can be considered constants inside the integral and, by the Equation (A.25), the Equation (B.36) yields:

$$\mathbf{M}_{lg} = \frac{1}{TV} \int_{t-T/2}^{t+T/2} \int_{A_{gl}(\mathbf{x},t)} \mathbf{n}_{lg} \cdot \left( -\tilde{p}_g \mathbf{I} + \tilde{\boldsymbol{\tau}}_g \right) dA \, dt'. \tag{B.37}$$

Therefore, by matching Equations (A.35) and (B.37), the relation between the interfacial force closures of both phases is the following:

$$\mathbf{M}_{lg} = -\mathbf{M}_{gl} \tag{B.38}$$

*B.5. The closure problem in terms of closure variables*

The set of Equations (B.10), (B.11), (B.23), (B.24) (B.25), (B.30) and (B.35) define the closure problem to be solved for momentum transport in gas-liquid bubbly flows. By considering that the length scale where the change of deviation variables is much smaller than the length scale where the averaged variables do it, it is possible to assume that the averaged variables are constants inside the averaging region. If the length scale for deviation variables characterizes the size of one representative and periodic region, then the subtracted system results in a problem which must be solved only for the deviation variables over such region. In addition, the average of deviation variables is zero and periodicity conditions at the external area can be imposed. Thus, we have that:

$$\langle \tilde{\mathbf{v}}_l \rangle^l, \langle \tilde{\mathbf{v}}_g \rangle^g, \langle \tilde{p}_l \rangle^l, \langle \tilde{p}_g \rangle^g = 0. \tag{B.39}$$

$$\tilde{\mathbf{v}}_k(\mathbf{x}'+\mathbf{l}_i,t') = \tilde{\mathbf{v}}_k(\mathbf{x}',t') \quad i=1,2,3, \quad k=l,g, \tag{B.40}$$

$$\tilde{p}_k(\mathbf{x}'+\mathbf{l}_i,t') = \tilde{p}_k(\mathbf{x}',t') \quad i=1,2,3, \quad k=l,g, \tag{B.41}$$

where $\mathbf{l}_i$ represent the lattice vectors which are required for describing a spatially periodic region. Because the scale constrains, the average variables can be assumed as constants and the closure problem can be expressed and solved in terms of closure variables [2,59]. The following linear solutions are proposed in terms of the non-homogeneous source:

$$\tilde{\mathbf{v}}_l = \mathbf{A}_l \cdot \left( \langle \mathbf{v}_g \rangle^g - \langle \mathbf{v}_l \rangle^l \right), \tag{B.42}$$

$$\tilde{p}_l = \mathbf{a}_l \cdot \left( \langle \mathbf{v}_g \rangle^g - \langle \mathbf{v}_l \rangle^l \right), \tag{B.43}$$

$$\tilde{\mathbf{v}}_g = \mathbf{A}_g \cdot \left( \langle \mathbf{v}_g \rangle^g - \langle \mathbf{v}_l \rangle^l \right), \tag{B.44}$$

$$\tilde{p}_g = \mathbf{a}_g \cdot \left( \langle \mathbf{v}_g \rangle^g - \langle \mathbf{v}_l \rangle^l \right). \tag{B.45}$$

where $\mathbf{A}_l$, $\mathbf{A}_g$, $\mathbf{a}_l$ and $\mathbf{a}_g$ are the closure variables which must be known in order to obtain the values for deviation variables. By substituting Equations (B.42)-(B.45) on the problem for the deviation variables, the system defined by the Equations (60)-(67) is obtained for the determination of the closure variables.